\documentclass[a4paper,11pt]{article}
\usepackage{jheppub}
\usepackage[T1]{fontenc}
\usepackage[english]{babel}
\usepackage{amsthm,amsfonts}
\usepackage[mathscr]{euscript}
\hyphenchar\font=\string"7F
\usepackage{tikz}
\usepackage{todonotes}
\usepackage{dsfont}
\usepackage{bbm}

\DeclareMathAlphabet{\mathdsl}{U}{bbm}{m}{sl}

\usepackage{amsfonts}

\newcommand{\dd}{\mathrm{d}}

\newcommand{\be}{\begin{equation}}
\newcommand{\ee}{\end{equation}}
\newcommand{\bea}{\setlength\arraycolsep{2pt} \begin{eqnarray}}
\newcommand{\eea}{\end{eqnarray}}
\newcommand{\nn}{\nonumber}

\def\ft#1#2{{\textstyle{\frac{\scriptstyle #1}{\scriptstyle #2} } }}
\def\fft#1#2{{\frac{#1}{#2}}}
\def\ep{{\epsilon}}
\def\del{{\partial}}
\def\Dslash{{\slashed D}}
\def\im{{\rm i}}
\def\dd{{\rm d}}

\newcommand{\bpm}{\begin{pmatrix}}
\newcommand{\epm}{\end{pmatrix}}

\newcommand{\Pb}{\overline{P}}
\newcommand{\DD}{\mathcal{D}}
\newcommand{\scT}{\mathcal{T}}

\def\ft#1#2{{\textstyle{\frac{\scriptstyle #1}{\scriptstyle #2} } }}
\def\fft#1#2{{\frac{#1}{#2}}}

\def\0{{\sst{(0)}}}
\def\1{{\sst{(1)}}}
\def\2{{\sst{(2)}}}
\def\3{{\sst{(3)}}}
\def\4{{\sst{(4)}}}
\def\5{{\sst{(5)}}}
\def\6{{\sst{(6)}}}
\def\7{{\sst{(7)}}}
\def\8{{\sst{(8)}}}
\def\sst#1{{\scriptscriptstyle #1}}
\def\oneone{\rlap 1\mkern4mu{\rm l}}
\def\del{{\partial}}

\def\im{{{\rm i\,}}}

\def\Dslash{\slash \negthinspace \negthinspace \negthinspace \negthinspace D}

\title{\boldmath Double Field Theory and Pseudo-Supersymmetry}
\preprint{MI-TH-2034}

\author[a]{Falk Hassler}
\author[a,b]{C.N. Pope}
\author[a]{and Hao-Yu Zhang}

\emailAdd{hassler@tamu.edu}
\emailAdd{pope@physics.tamu.edu}
\emailAdd{haoyuzhang001@gmail.com}

\affiliation[a]{George P. \& Cynthia Woods Mitchell Institute for Fundamental Physics and Astronomy,\\ Texas A\&M University, College Station, TX 77843, USA}
\affiliation[b]{DAMTP, Centre for Mathematical Sciences,\\Cambridge University,
Wilberforce Road, Cambridge CB3 OWA, UK}

\abstract{
Many of the useful features of supergravities, such as admitting
supersymmetric bosonic backgrounds governed by first-order BPS equations, can be realised in a much broader setting by relaxing the requirement of closure of the superalgebra beyond the level of quadratic fermion terms. The resulting pseudo-supersymmetric theories can be defined in arbitrary spacetime dimensions.  We focus here on the ${\cal N}=1$ pseudo-supersymmetric extensions of the arbitrary-dimensional bosonic string action, which were constructed a few years ago. In this paper, we recast these in the language of generalised geometry. More precisely, we construct the action and the corresponding supersymmetry transformation rules in terms of O($D$)$\times$O($D$) covariant derivatives, and we discuss consistent truncations on manifolds with generalised $G$-structure. As explicit examples, we discuss Minkowski$\times G$ vacuum solutions and their corresponding pseudo-supersymmetry. We also briefly discuss squashed group manifold solutions, including an example with a Lorentzian signature metric on the group manifold $G$.} 

\begin{document}

\maketitle

\section{Introduction}
Supersymmetry provides a powerful tool for probing aspects of physics that would otherwise be beyond the limits of computability. One important example is that the second-order non-linear field equations of Einstein gravity or supergravity can be reduced to first-order equations in certain circumstances, namely when there exist supersymmetric bosonic backgrounds that admit one or more Killing spinors. Beyond the classical level, supersymmetry severely restricts quantum corrections and allows some non-perturbative results to be obtained.

A feature of supersymmetry is that it implies restrictions on the dimension of the spacetime. In particular, beyond 11 dimensions it is not possible to find any supersymmetric extension of gravity without adding higher spin fields, and this rules out having a Lagrangian description. One might therefore conclude that supersymmetry would in general be of no help in the study of theories in arbitrary higher dimensions. However, these theories can still possess a pseudo-supersymmetry \cite{Lu:2011zx,Liu:2011ve,Lu:2011nga,Lu:2011ku}, which is a weaker notion of supersymmetry that only involves fermionic terms up to second order, in the action and the transformation rules. This is in fact sufficient in order to be able to derive many of the useful features of conventional supersymmetric theories, including the existence of pseudo-Killing spinors in certain backgrounds. Thus pseudo-supersymmetry still allows the second-order field equations for the bosonic fields to be reduced into first-order Bogomol'nyi–Prasad–Sommerfield (BPS) conditions in such backgrounds. Hence, it can provide a powerful tool in the study of solutions of theories of gravity coupled to matter in arbitrary dimensions. 

A well-established framework for exploring the landscape of supergravity vacua is provided by (exceptional) generalised geometry \cite{Hitchin:2004ut,Gualtieri:2003dx,Hull:2007zu,Coimbra:2011nw,Coimbra:2011ky,Coimbra:2012af}, and the closely related double or exceptional field theory (DFT/ExFT)\ \cite{Siegel:1993xq,Siegel:1993th,Hull:2009mi,Hohm:2010jy,Hohm:2010pp,Jeon:2010rw,Jeon:2011cn,Berman:2010is,Berman:2011cg,Berman:2011pe,Berman:2012vc,Cederwall:2013naa,Hohm:2013pua,Hohm:2013vpa,Hohm:2013uia,Hohm:2014fxa,Musaev:2014lna,Ciceri:2016dmd}\footnote{There has been a considerable amount of original work in this field and therefore we only reference a few key contributions here, which is of course highly subjective. We refer to the reviews \cite{Koerber:2010bx,Aldazabal:2013sca,Hohm:2013bwa,Berman:2020tqn} for a complete list of references.}. Particularly interesting for our work are the supersymmetric extensions of bosonic DFT \cite{Hohm:2011ex,Hohm:2011nu,Jeon:2011sq,Hohm:2011zr,Hohm:2011dv,Jeon:2012kd,Jeon:2011vx,Jeon:2012hp}. All these approaches share one defining property, namely the unification of local diffeomorphisms with form-field gauge transformations into one unified symmetry group. In the most basic setup, the metric and the two-form $B$-field potential of a bosonic string action are combined into the generalised metric, giving rise to the O($n$,$n$) symmetry of DFT. Equivalently, this structure is captured by the generalised tangent bundle $T M \oplus T^* M$ of generalised geometry. Note that in general, DFT is capable to capture backgrounds which go beyond supergravity and generalised geometry \cite{Dibitetto:2012rk,Geissbuhler:2013uka}. Here, however, we shall be concerned with the most conservative case, where the section condition is satisfied globally. In this case DFT is just a rewriting of supergravity, and it is completely equivalent to generalised geometry. Still, there are two important advantages of this rewriting: First, supersymmetry variations have a much simpler form and second, abelian T-duality becomes a manifest symmetry of the string's low-energy effective target space action. A natural question in this context is whether pseudo-supersymmetry permits a similar treatment. We answer this question in the affirmative, and demonstrate that it is possible to extend the existing results of supersymmetric DFT from ten dimensions to arbitrary dimensions.

An important application is the construction of pseudo-supergravity vacua. In particular, we combine the technique of consistent truncations with pseudo-supersymmetry to show how the field equations in arbitrary dimensions can be simplified significantly. More precisely, in the examples we consider, the consistent truncation renders the field equations algebraic but still quadratic. Clearly this is already a major simplification, but still quadratic equations with multiple variables can be hard to solve. A similar problem arises in the classification of Lie algebras, whose Jacobi identity is a quadratic constraint. In low dimensions, it is possible to solve it, and this gives rise to a complete classifications of real Lie algebra up to six dimensions \cite{Snobl2014}. Beyond that, solving the quadratic constraint becomes forbiddingly complicated. A similar situation is encountered in the prototypical example of consistent truncations in DFT, namely in generalised Scherk-Schwarz reductions \cite{Geissbuhler:2011mx,Aldazabal:2011nj,Grana:2012rr,Berman:2012uy,Berman:2013cli,Hassler:2014sba}. Compared to a standard geometric reduction on a group manifold with isometry group $G_L \times G_R$, which retains the singlets under either $G_L$ or $G_R$, the consistent reductions in DFT allow one to retain all the gauge bosons of the complete isometry group. This is a much more complicated reduction, because of the potentially dangerous trilinear coupling of massive spin-2 modes to bilinears constructed from the $G_L\times G_R$ Yang-Mills bosons \cite{Duff:1984hn}. The existence of a consistent reduction of the $(n+D)$ bosonic string to a $D$-dimensional group manifold keeping all of the $G_L\times G_R$ gauge bosons was conjectured in \cite{Duff:1986ya}, with further supporting evidence found in \cite{Cvetic:2003jy}. A complete proof of the consistency was obtained in \cite{Baguet:2015iou}, utilizing the $O(D,D)$ formulation of $(n+D)$-dimensional bosonic string \cite{Hohm:2013nja}. Combining a generalised Scherk-Schwarz reduction with pseudo-supersymmetry, we show how the quadratic field equations for the remaining fields can be reduced, in appropriate backgrounds, to linear equations. Because of the less restrictive nature of pseudo-supersymmetry, in comparison to ordinary supersymmetry, this can be done in arbitrary spacetime dimensions.

The paper is organized as follows. In section 2, we give a short review of the $\mathcal{N}=1$ pseudo-supersymmetric theory. In section 3, we reformulate it in terms of generalised geometry and then spell out the conditions for the existence of a consistent truncation. In sections 4 we explicitly construct solutions of the form  (Minkowski)$_{D-\dim G}\times G$, including a description in the framework of generalised geometry. In section 5 we discuss their pseudo-supersymmetry, both in standard field theory and in generalised geometry. In an appendix, we construct an example of a (Minkowski)$_{D-\dim G}\times G$ background, for $G=SO(5)$, where the metric on the group is squashed.  It turns out to have Lorentzian signature.

\section{Pseudo-Supersymmetrised Bosonic String}\label{sec:lagtran}
As described in \cite{Lu:2011zx} one can construct a pseudo-supersymmetric fermionic extension of the bosonic string Lagrangian, in a completely arbitrary dimension $D$. That is to say, there exist supersymmetry-like transformation rules that leave the Lagrangian invariant, modulo terms beyond the quadratic order in fermions. In practice many of the desirable features of supersymmetry, such as the existence of Killing spinors in bosonic backgrounds, BPS conditions and first-order equations, do not directly depend upon the full closure of the transformations. This means that all the useful consequences of having fermionic symmetries in bosonic backgrounds will equally well arise in the much larger arena of pseudo-supersymmetric theories.

\subsection{Lagrangian and pseudo-supersymmetry transformation rules}
The Lagrangian for the pseudo-supersymmetric extension of the bosonic string in an arbitrary dimension was constructed in \cite{Lu:2011zx}, where it was presented both in the Einstein frame and in the string frame. Here, we reproduce the result from \cite{Lu:2011zx} in the string frame, with the following notational changes. Firstly, we denote the spacetime dimension by $D$ rather than $d$, since in this paper $d$ will be reserved to denote the generalised dilaton of DFT. Secondly, in order to harmonise our notation with some of the DFT literature, we perform the rescalings
\be
  \psi_\mu\rightarrow \sqrt{2}\, \psi_\mu\,,\qquad \lambda
  \rightarrow \sqrt{2}\,\lambda\,,\qquad
  \ep \rightarrow \sqrt{2}\,\ep\,, \label{rescale}
\ee
on the fermion fields and pseudo-supersymmetry parameter, and finally, we make the replacement $\Gamma_a\rightarrow -\Gamma_a$, which of course preserves the Clifford algebra. With these replacements, the $D$-dimensional pseudo-supersymmetric Lagrangian of \cite{Lu:2011zx}, in the string frame, becomes
\bea
  e^{-1} {\cal L} &=& e^{-2\Phi}\, \Big[ R +
  4(\del\Phi)^2 - \ft1{12} H^2 -
  \bar\psi_\mu\Gamma^{\mu\nu\rho}D_\nu\psi_\rho +
  \bar\lambda \Dslash\lambda -2 \im\,\sqrt{\beta}\,
  \bar\lambda\Gamma^{\mu\nu}\, D_\mu \psi_\nu
  \nn\\
  &&\qquad-2 \bar\psi_\mu \Gamma^\mu\psi_\rho \,\del^\rho\Phi +
   \fft{2 \im}{\sqrt{\beta}}\,
  \bar\psi_\mu \Gamma^\nu\Gamma^\mu\lambda \,
          \del_\nu\Phi\label{eqn:LpseudoSUSY}\\
  && +  H_{\nu\rho\sigma} \Big\{\ft1{24}
  \bar\psi_\mu\Gamma^{\mu\nu\rho\sigma\lambda}\, \psi_\lambda + \ft14
    \bar\psi^\nu\Gamma^\rho\psi^\sigma - \ft1{24}\,
    \bar\lambda\Gamma^{\nu\rho\sigma}\lambda+ \ft{\im}{12\sqrt{\beta}}\,
  \bar\psi_\mu \Gamma^{\mu\nu\rho\sigma} \lambda \Big\} \Big]\,,\nn
\eea
and the pseudo-supersymmetry transformation rules are given by
\bea 
  \delta\psi_\mu &=& D_\mu\ep -\ft18  H_{\mu\nu\rho}\,
    \Gamma^{\nu\rho}\, \ep
     \,,\nn\\
  \delta\lambda &=&
    \im \sqrt{\beta}\Big(\Gamma^\mu \partial_\mu \Phi -
    \ft1{12} \Gamma^{\mu\nu\rho} H_{\mu\nu\rho}\Big)\ep\,,\nn\\
  \delta e^a_\mu &=& -\ft12 \bar\psi_\mu \Gamma^a\, \ep\,,\nn\\
  \delta\Phi &=& -\fft{\im}{4\sqrt{\beta}}\, \bar\ep\, \lambda\,,\nn\\
  \delta B_{\mu\nu} &=& \bar\ep \Gamma_{[\mu} \psi_{\nu]} \,.
  \label{eqn:SUSYtrafo1}
\eea
Note that $\delta\psi_\mu$ may be re-expressed in terms of a torsionful connection as $\delta\psi_\mu = D_\mu(\omega_-)\ep$, where
\be 
  \omega_{\mu\pm}^{ab} \equiv \omega_\mu^{ab} \pm \ft12 
H_\mu{}^{ab}\,. 
\ee

The constant $\beta$, which is either $+1$ or $-1$ depending on the dimension $D$ and the spinor representation, characterises the symmetry property of the gamma matrices,
\be\label{eqn:defbeta}
  \Gamma_\mu^T=\beta C\Gamma_\mu C^{-1}\,.
\ee
It is listed for each dimension and representation in table~\ref{tab:Spinors} in the appendix \ref{app:Spinors}. Many further properties of spinors in diverse dimensions are summarised in our notation in \cite{Lu:2011zx}. All coefficients in (\ref{eqn:LpseudoSUSY}) and (\ref{eqn:SUSYtrafo1}) were determined by the requirement that the Lagrangian be invariant under the pseudo-supersymmetry transformations, provided that one neglects fermionic terms that would arise from higher fermionic powers in the Lagrangian or pseudo-supersymmetry transformations. 

It was shown in \cite{Lu:2011ku} that, just like in the case of the supersymmetry transformations for ten-dimensional ${\cal N}=1$ supergravity, the integrability conditions obtained by taking commutators of the pseudo-supersymmetry transformations on a bosonic background are satisfied if the full set of field equations for the $D$-dimensional bosonic string are satisfied. 

\subsection{Adding a conformal anomaly term}
As was shown in \cite{Lu:2011ku}, one can also add a ``conformal anomaly'' term to the Lagrangian. In the string frame, after performing the rescalings \eqref{rescale} and the replacement $\Gamma_a\rightarrow -\Gamma_a$ detailed above, the additional terms in the Lagrangian take the form
%%%%%
\be
  e^{-1} {\cal L}_{c} =  e^{-2\Phi}\, \Big[
  -\frac{m^2}{2}-\frac{m}{2\sqrt{2\beta}} \Big (\bar \psi_\mu \Gamma^{\mu\nu}\psi_\nu+2\sqrt{-\beta}\bar \psi_\mu \Gamma^\mu \lambda-\bar \lambda\lambda \Big) \Big]\,.\label{conflag}
\ee
%%%%%
There are associated additional terms in the fermion transformation rules, given by
%%%%%
\begin{equation}
  \delta_{\rm extra} \psi_\mu =0\,,\qquad
  \delta_{\rm extra} \lambda = \ft{\im}{2\sqrt{2}} m\, \ep\,.
  \label{conftrans}
\end{equation}
%%%%%
Note that the fermionic extension of the conformal anomaly term in \eqref{conflag} really requires a doubling of the fermionic degrees of freedom. This is most easily stated in dimensions $D= 2\,\, \hbox{mod}\, 8$, where we can choose $\beta=-1$ and the basic spinors of the pseudo-supersymmetrised bosonic string would be both Majorana and Weyl (with $\psi_\mu$ and $\ep$ being chiral, and $\lambda$ anti-chiral). The fermionic terms in \eqref{conflag} would vanish under these conditions, but will be non-vanishing if the chirality constraints on the fermions are removed. In cases where $\beta=+1$, the first two fermionic terms in \eqref{conflag} will vanish identically, if the spinors are Majorana or symplectic-Majorana.  In these cases, one can still pseudo-supersymmetrise the conformal anomaly term if one doubles the number of fermions, by adding an additional doublet index,
%%%%%
\be
  \psi_\mu\longrightarrow \psi_\mu^\alpha\,,\qquad
  \lambda\longrightarrow \lambda^\alpha\,.
\ee
%%%%%
All the previous fermion bilinears in the Lagrangian will now have $\alpha$ and $\beta$ indices contracted with $\delta_{\alpha\beta}$. The terms in ${\cal L}_c$, on the other hand, will have the $\alpha$ and $\beta$ indices contracted with $\ep_{\alpha\beta}$. An $\ep_{\alpha\beta}$ should also be inserted in the extra terms (\ref{conftrans}) in transformation rules for $\psi_\mu$ and $\lambda$.

\section{Generalised Geometry and Pseudo-Supersymmetry}\label{sec:genpseudo}
It is possible to simplify the Lagrangian \eqref{eqn:LpseudoSUSY} considerably by introducing the generalised dilaton
%%%%%
\begin{equation}\label{eqn:gendil}
  d = \Phi - \frac12 \log e\,,
\end{equation}
%%%%%
and its superpartner
%%%%%
\begin{equation}
  \rho = \Gamma^\mu\, \psi_\mu + \frac{\im}{\sqrt{\beta}} \lambda\,.
\end{equation}
%%%%%
Furthermore, we unify the frame field and the $B$-field by introducing the generalised frame field with the components
%%%%%
\begin{equation}
  \begin{aligned}
    E^{(+)}_a &= \frac1{\sqrt{2}} \left( e_a^\mu \partial_\mu + e_{\mu a} 
      \dd x^\mu - \iota_{e_a} B \right)\,, \\
    E^{(-)}_a &= \frac1{\sqrt{2}} \left( e_a^\mu \partial_\mu - e_{\mu a}  
      \dd x^\mu - \iota_{e_a} B \right)\,,
  \end{aligned}
\end{equation}
%%%%%
and $\iota_{e_a} B = e_a^\mu B_{\mu\nu} \dd x^\nu$. Each of these $2 D$ components is a generalised vector on the generalised tangent space $T M \oplus T^* M$. After this identification and the redefinitions above, the pseudo-supersymmetry transformation rules 
\eqref{eqn:SUSYtrafo1} and \eqref{conftrans} can be written in the compact form
%%%%%
\begin{equation}\label{eqn:susyvarGG}
  \begin{aligned}
    \delta \psi_\mu &= \nabla^{(-)}_\mu \ep\,, & 
      \delta_{\rm extra} \psi_\mu &= 0\,, \\
    \delta \rho &= \Gamma^\mu\, \nabla^{(+)}_\mu \ep\,, & \qquad\qquad
      \delta_{\rm extra} \rho &= - \frac1{2\sqrt{2\beta}} m \ep\,, \\
      \langle E^{(-)}_b , \delta E^{(+)}_a \rangle &= -\frac12 \bar \ep \Gamma_b \psi_a \,, \\
    \delta d &= - \frac14 \bar \ep\, \rho\,,
  \end{aligned}
\end{equation}
%%%%%
where the O($D$)$\times$O($D$) covariant derivatives $\nabla^{(\pm)}_\mu$ play a crucial role.  They are defined by \cite{Coimbra:2011nw,Hohm:2011nu}
%%%%%
\begin{equation}
  \begin{aligned}
    \nabla^{(-)}_\mu \ep &= \left( D_\mu - \ft18 H_{\mu\nu\rho} 
      \Gamma^{\mu\nu} \right) \ep \,, \\
    \Gamma^\mu \nabla^{(+)}_\mu \ep &= 
\left( \Gamma^\mu D_\mu - \ft1{24} H_{\mu\nu\rho} 
\Gamma^{\mu\nu\rho} - \Gamma^\mu \del_\nu \Phi \right) \ep\,,
  \end{aligned}
\end{equation}
%%%%%
and as we will see in the next subsection, they also have very nice properties when it comes to consistent truncations.

Like the pseudo-supersymmetry transformation rules, also the action \eqref{eqn:LpseudoSUSY} simplifies considerably once written in string frame and after applying the redefined fields and the adapted covariant derivatives,
\begin{equation}
  e^{2 d} \mathcal L_D = R + 4 (\partial \phi)^2 - \frac1{12} H^2_{(3)} - \bar\psi^a \Gamma^b \nabla_b^{(+)} \psi_a -\beta \bar\rho \Gamma^a \nabla_a^{(+)} \rho + 2 \bar\psi^a \nabla_a^{(-)} \rho \,,
\end{equation}
where $\beta$ is defined by the use of charge conjugation matrix $C$ in \eqref{eqn:defbeta}. 

In ten dimensions, this Lagrangian matches the one of ${\cal N}$=1 Double Field Theory \cite{Hohm:2011nu} after implementing the solution of the section condition which removes the dependence on the coordinates conjugate to string winding modes and choosing the parameter $\beta=-1$. The interesting observation here is that this result even holds in arbitrary dimensions once we drop the additional constraints imposed by supersymmetry in favour of pseudo-supersymmetry.

The conformal anomaly terms \eqref{conflag} can also be formulated in the language of generalised geometry,
\be
  e^{2d} {\cal L}_{c} = -\frac{m^2}{2} - \frac{m\sqrt{\beta}}{2\sqrt{2}} \left(\bar \rho \rho -\beta \bar \psi_\mu \psi^\mu\right)\,.
\ee
To satisfy the pseudo-supersymmetry property, it follows from \eqref{conftrans} that the variation rule of 
$\rho$ needs to be modified by $\delta_{\rm extra}$ in \eqref{eqn:susyvarGG}.

\subsection{Consistent truncations}\label{sec:consistenttr}
The crucial observation for constructing consistent truncations in the $\mathcal{N}=1$ pseudo-supersymmetric theory is that all relevant quantities like the Lagrangian, the pseudo-supersymmetry transformation rules and the field equations can be written in terms of covariant derivatives $\nabla^{(\pm)}_\mu$. To construct them, one starts with an O($D$,$D$) structure which is defined by the invariant metric
\begin{equation}
  \eta_{AB} = \begin{pmatrix} \eta_{ab} & 0 \\ 0 & -\eta_{\bar a\bar b} \end{pmatrix}
    \quad \text{and} \quad
  \eta^{AB} = \begin{pmatrix} \eta^{ab} & 0 \\ 0 & -\eta^{\bar a\bar b} \end{pmatrix}\,,
\end{equation}
($\eta_{ab} = \eta_{\bar a\bar b}$ is the invariant metric of O($D$) or its Lorentzian counterpart) which is also used to raise and lower ``doubled'' indices $A$, $B$, \ldots. The group O($D$,$D$) can further be broken to O($D$)$\times$O($D$) by requiring a second invariant metric, the generalised metric,
\begin{equation}
  \mathcal{H}_{AB} = \begin{pmatrix} \eta_{ab} & 0 \\ 0 & \eta_{\bar a\bar b} \end{pmatrix}
    \quad \text{and} \quad
    \mathcal{H}^{AB} = \begin{pmatrix} \eta^{ab} & 0 \\ 0 & \eta^{\bar a\bar b} \end{pmatrix}\,.\label{HAB}
\end{equation}
It encodes the metric and the $B$-field once it is pulled to the generalised tangent space $T M \oplus T^*M$, where it reads
\begin{equation}\label{eqn:genmetric}
  \mathcal{H}^{IJ} = \begin{pmatrix} g_{ij} - B_{ik} g^{kl} B_{lj} & \quad-B_{ik} g^{kj} \\
    g^{ik} B_{kj} & g^{ij}
  \end{pmatrix}\,.
\end{equation}
Note that we here have switched from using Greek indices, $\mu$, $\nu$, \ldots, for spacetime coordinates to Latin indices $i$, $j$, \ldots. We do this because in DFT, there is a need for capital, doubled, indices as well as lower case, standard, indices, and the Greek alphabet does not lend itself to this distinction.

The metrics (\ref{HAB}) and (\ref{eqn:genmetric}) are related by the generalised frame,
\begin{equation}
  \mathcal{H}^{IJ} = E_A{}^I E_B{}^J \mathcal{H}^{AB}\,, \qquad \text{with} \qquad
  E_A{}^I \begin{pmatrix} \dd x^i \\ \partial_i \end{pmatrix} = \begin{pmatrix} E_a^{(+)} \\ E_{\bar a}^{(-)}  \end{pmatrix}\,.
\end{equation}

In order to construct consistent truncations, one restricts the form of the covariant derivative $\nabla_A = \ft{1}{\sqrt{2}}( \nabla^{(+)}_a , \nabla^{(-)}_{\bar a} )$ to
\begin{equation}\label{eqn:defnabla}
  \nabla_A V^B = \DD_A V^B + \omega_{AC}{}^B V^C\,,
\end{equation}
where $\DD_A$ is a second covariant derivative which admits some invariant tensors and thus defines a generalised $G$-structure \cite{ Cassani:2019vcl}. $G$ can be any subgroup of O($D$)$\times$O($D$); we present some examples later but for the moment we keep the discussion general. To obtain $\nabla_A$ from $\DD_A$ the tensor $\omega_{AB}{}^C$ has to be fixed. This is done by imposing four constraints on $\nabla_A$ \cite{Hohm:2011si}:
\begin{enumerate}
  \item It is compatible with the O($D$,$D$) metric:
    \begin{equation}
      \nabla_A \eta_{BC} = 0\,,
        \quad \text{which implies} \quad
        \omega_{ABC} = - \omega_{ACB}\,.
    \end{equation}
  \item It is compatible with the generalised metric:
    \begin{equation}
      \nabla_A \mathcal{H}_{BC} = 0\,.
    \end{equation}
  \item It is compatible with integration by parts:
    \begin{equation}\label{eqn:defFA}
      \int e^{-2 d} \nabla_A V^A = 0\,,
        \quad \text{which implies} \quad
      \omega^B{}_{BA} = 2 \DD_A d  - \partial_I E_A{}^I := F_A\,.
    \end{equation}
  \item It has vanishing generalised torsion, implying
    \begin{equation}\label{eqn:scT=0}
      3 \omega_{[ABC]} = - \scT_{ABC} := F_{ABC}\,,
    \end{equation}
    where $\scT_{ABC}$ is the generalised torsion of $\DD_A$.
\end{enumerate}
These constraint do not fix $\omega_{AB}{}^C$ completely. However, all physically relevant quantities like the action, the field equations and the pseudo-supersymmetry transformations use $\nabla^{(\pm)}_\mu$ in such a way that the undefined contributions drop out. To present the partially fixed $\omega_{AB}{}^C$, it is convenient to introduce the projectors
\begin{equation}
  P_{AB} = \frac12 ( \eta_{AB} + \mathcal{H}_{AB} ) \quad \text{and} \quad
  \Pb_{AB} = \frac12 ( \eta_{AB} - \mathcal{H}_{AB} )\,,
\end{equation}
which project onto the two factors of O($D$)$\times$O($D$),
\begin{equation}
  P^A{}_B = \begin{pmatrix} \delta^a_b & 0 \\ 0 & 0 \end{pmatrix}
    \quad \text{and} \quad
  \Pb^A{}_B = \begin{pmatrix} 0 & 0 \\ 0 & \delta^{\bar a}_{\bar b} \end{pmatrix} \,.
\end{equation}
The constraints above restrict the form of $\omega_{ABC}$ to
\begin{equation}\label{eqn:formomega}
  \omega_{ABC} = \left( \frac13 P^D{}_A P^E{}_B P^F{}_C + \Pb^D{}_A P^E{}_B P^F{}_C + P^D{}_A \Pb^E{}_B \Pb^F{}_C + \frac13 \Pb^D{}_A \Pb^E{}_B \Pb^F{}_C \right) X_{DEF} 
\end{equation}
with
\begin{equation}
  X_{ABC} = F_{ABC} + \frac{6}{D-1} \eta_{A[B} F_{C]} \,.
\end{equation}
Note that there are still unconstrained components of $\omega_{ABC}$ left \cite{Hohm:2011si}. But these are irrelevant in the calculation of all physically relevant quantities and thus we can safely ignore them. A consistent truncation arises if the action of $\nabla_A$ on a tensor invariant under $D_A$ gives rise to another (or the same) tensor that is invariant under $\DD_A$. In this case, the set all these invariant tensors forms a consistent truncation. By using the definition of $\nabla_A$ in \eqref{eqn:defnabla}, this requirement translates to the two constraints
\begin{equation}
  \DD_A F_{BCD} = 0 \quad \text{and} \quad \DD_A F_B = 0\,. \label{coneq}
\end{equation}

\section{Minkowski\texorpdfstring{$\times G$}{xG} Group Manifold Compactifications}
It was observed in \cite{Duff:1986ya} that the $d$-dimensional bosonic string with the added conformal anomaly term admits a vacuum solution of the form (Minkowski)$_{D-\dim G}\times G$, where $G$ is any semi-simple compact $\dim G$-dimensional Lie group. Here, we shall study the pseudo-supersymmetry of these vacuum solutions. In order to do this, we first need to establish some basic notation and results for group manifold compactifications.

\subsection{Conventions and geometry for group manifolds}\label{sec:conventions}
The vacuum solution employs the group manifold $G$ equipped with its bi-invariant metric $g_{mn}$. This has left-acting and right-acting Killing vectors of the group $G$, which we denote by $K_{L\, a}^m$ and $K_{R\, a}^m$ respectively. They obey the algebra
%%%%%
\be
  [K_{L\, a},K_{L\, b}]=-c\,f_{ab}{}^c\, K_{L\, c}\,,\qquad
  [K_{R\, a},K_{R\, b}]=c\, f_{ab}{}^c\, K_{R\, c}\,,\qquad 
  [K_{L\, a},K_{R\, b}]=0\,,\label{Kcoms}
\ee
%%%%%
where $f_{ab}{}^c$ are the structure constants, and $c$ is a scale-setting constant. The Killing vectors may be normalised so that
%%%%%
\be
  g_{mn}\, K_{L\, a}^m\, K_{L\, b}^n=\delta_{ab}\,,\qquad
  g_{mn}\, K_{R\, a}^m\, K_{R\, b}^n=\delta_{ab}\,,
\ee
%%%%%
with $\delta_{ab}$ being proportional to the Cartan-Killing metric,
%%%%%
\be
  -f_{ac}{}^d \, f_{bd}{}^c= C_A\, \delta_{ab}\,,
\ee
%%%%%
where $C_A$ is the quadratic Casimir of the group $G$. Conversely, one has
%%%%%
\be
  g^{mn}= K_{L\, a}^m\, K_{L\, b}^n\, \delta^{ab}= 
    K_{R\, a}^m\, K_{R\, b}^n\, \delta^{ab}\,.
\ee
%%%%%
It follows that one may view either the $K_{L\, a}^m$ or the $K_{R\, a}^m$ Killing vectors as defining a vielbein $e^a=e^a_m\, dy^m$. We shall consider the left-invariant vielbein
%%%%%
\be
  e^a= K_{R}^a=  K_{R\, m}^a\, dy^m\,.\label{viel}
\ee
%%%%%

Using \eqref{Kcoms}, the 1-forms $K_{L}^a$ and $K_{R}^a$ obey
%%%%%
\be
  dK_L^a= \ft12 c\, f_{bc}{}^a\, K_L^b\wedge K_L^c\,,\qquad
  dK_R^a= -\ft12 c\, f_{bc}{}^a\, K_R^b\wedge K_R^c\,,
\ee
%%%%%
The vielbein \eqref{viel} therefore obeys $de^a=-\ft12 c\, f_{bc}{}^a\, e^b\wedge e^c$, and so the torsion-free spin-connection, defined by $de^a=-\omega^a{}_b\, \wedge e^b$ and $\omega_{ab}=-\omega_{ba}$ is therefore given by
%%%%%
\be
  \omega_{ab}= -\ft12 c\, f_{abc}\, e^c\,.\label{spincon}
\ee
%%%%%
Note that since we are taking $G$ to be compact and semi-simple, $f_{abc}$ is totally antisymmetric. The curvature 2-forms $\Theta_{ab}=d\omega_{ab} + \omega_{ac}\wedge \omega_{cb}$ and the Riemann tensor (following
from $\Theta_{ab}=\ft12 R_{abcd}\, e^c\wedge e^d$) are then given by\footnote{One needs to use the Jacobi identity to show this.}
%%%%%
\be
  \Theta_{ab}=\ft18 c^2\, f_{abe}\, f_{cde}\, e^c\wedge e^d\,,
  \qquad
  R_{abcd}=\ft14 c^2\,  f_{abe}\, f_{cde}\,.
\ee
%%%%%
Finally, we have the Ricci tensor and Ricci scalar, given by
%%%%%
\be
  R_{ab}= \ft14 c^2\,C_A\, \delta_{ab}\,,\qquad
  R=\ft14 c^2\, C_A\, \dim G\,.
\ee
%%%%%

Note that $f_{abc}$ is covariantly constant.  The Lorentz-covariant exterior derivative $D$ acts on Lorentz vector as $DV^a=dV^a +\omega^a{}_b\, V^b$, so
%%%%%
\be
Df^{abc}= df^{abc} + \omega^{[a}{}_d\, f^{bc]d}\,,
\ee
%%%%%
and since the $f^{abc}$ are constants, and 
the spin connection is given by (\ref{spincon}), we have
%%%%%
\be
D f^{abc}= -\ft12 c f^{[a}{}_{de}\, f^{bc] d}\, K^e_R \,,
\ee
%%%%%
and this vanishes by virtue of the Jacobi identity.  Thus it follows
that in coordinate indices we also have $\nabla_m\, f_{npq}=0$.

\subsection{Consistent truncations}
Following the discussion in section~\ref{sec:consistenttr}, we now construct a consistent truncation on the group manifolds $G$. To do so, our first objective is to fix an appropriate covariant derivative $\DD_A$. We impose now, that $\DD_A$ annihilates the generalised frame field $E_B{}^I$. This renders the corresponding generalised geometry parallelisable or, equally, the generalised structure group trivial. More specifically, we have
\begin{equation}
  \DD_I E_A{}^J = \partial_I E_A{}^J + \Gamma_{IK}{}^J E_A{}^K = 0\,,
\end{equation}
thus determining the corresponding connection
\begin{equation}
  \Gamma_{IJK} = \partial_I E^A{}_J E_{AK}\,.
\end{equation}
Its generalised torsion is given by
\begin{equation}
  \scT_{IJK} = 3 \Gamma_{[IJK]}\,,
\end{equation}
and therefore we obtain from \eqref{eqn:scT=0}
\begin{equation}\label{eqn:gentorsion}
  F_{ABC} = 3 E_{[A}{}^I \partial_I E_B{}^J E_{C]J} \,.
\end{equation}
A consistent truncation requires that \eqref{coneq} hold. Thus, we have to find a generalised frame field $E_A{}^I$ on the group manifold $G$ such that
\begin{equation}\label{eqn:constF}
  \DD_I F_{ABC} = \partial_I F_{ABC} = 0 
\end{equation}
holds. Equivalently stated, the generalised torsion in flat indices 
must be constant.

This problem does not have a unique solution, because there are an infinite number of admissible generalised frame fields that satisfy \eqref{eqn:constF} on the Lie group $G$. For definiteness, we choose here the solution discussed in \cite{Baguet:2015iou}. It is given by
\begin{equation}
  \begin{aligned}
    \sqrt{2} E^{(+)}_a &= K_{L a}^m \partial_m - \eta_{ab} \left(\iota_{K_L^b} B - K_{L m}^b \dd x^m \right)\,, \\
    \sqrt{2} E^{(-)}_a &= K_{R a}^m \partial_m - \eta_{ab} \left(\iota_{K_R^b} B + K_{R m}^b \dd x^m \right)\,,
  \end{aligned}
\end{equation}
where $K_R$ and $K_L$ denote the left- and right-invariant vectors fields and their respective duals from subsection \ref{sec:conventions}, and $\iota_X \,B=X^m\, B_{mn}\, dx^n$ for any vector $X$ . Additionally, we also have to incorporate a $B$-field whose corresponding $H$-flux yields
\begin{equation}
  \dd B = - \frac{c}{3!} f_{abc} K^a_R \wedge K^b_R \wedge K^c_R\,. 
\end{equation}
For this generalised frame field, we now compute the generalised torsion \eqref{eqn:gentorsion} with the non-vanishing components
\begin{equation}\label{eqn:FsforG}
  F_{abc} = \frac{c}{\sqrt{2}} f_{abc} \qquad \text{and} \qquad
  F_{\bar a\bar b\bar c} = \frac{c}{\sqrt{2}} f_{\bar a\bar b\bar c}\,.
\end{equation}
Note that $f_{abc} = f_{ab}{}^d \delta_{dc} = f_{\bar a\bar b\bar c}$ coincides with the structure coefficients that govern the generators of the Lie group $G$. They appear here because of the Killing vectors algebra \eqref{Kcoms}.

We also need to compute the flux $F_A$, which captures the dilaton, and check that it is constant, as required by the second equation in \eqref{coneq}. By combining \eqref{eqn:gendil} with \eqref{eqn:defFA}, we obtain
\begin{equation}
  F_A = 2 \DD_A \phi - \DD_A \log \det e - \partial_I E_A{}^I\,.
\end{equation}
This equation splits into two contributions, for $F_a$ and $F_{\bar a}$ respectively. Let us take a closer look at
\begin{equation}
  F_a = 2 K_{L a}^m \partial_m \phi - K_{L a}^m \partial_m K_{L n}^b K_{L b}^n - \partial_m K_{L a}^m\,,
\end{equation}
where we take into account that $e^a_m$ can be identified with $K_{L m}^a$. The right-hand side of this relation can be further simplified by using \eqref{Kcoms}, yielding
\begin{equation}
  F_a = 2 K_{L a}^m \partial_m \phi - c f_{ab}{}^b \,.
\end{equation}
The last term vanishes because we take $G$ to be semi-simple. An analogous argument applies to $F_{\bar a}$. Hence, we conclude that $F_A$=const. requires a linear dilaton. Finally, the Bianchi identity for $D_I$ implies that
\begin{equation}\label{eqn:biFA}
  F_{AB}{}^C F_C = 0
\end{equation}
must hold. Since the generalised torsion $F_{ABC}$ matches the structure coefficients of the isometry group $G_L\times G_R$, $F_C$ is in one-to-one correspondence with an element in the center of this group. But because $G$ is semi-simple, so is $G_L\times G_R$. Semisimple Lie groups have a trivial center, and therefore only $F_A=0$ is consistent with the Bianchi identity \eqref{eqn:biFA}. Thus we conclude that the dilaton must be constant in order to give rise to a consistent truncation.

\subsection{The Minkowski\texorpdfstring{$\times G$}{xG} vacuum}
At this point, it is convenient to change the index labelling conventions and notation a little, and rewrite the Lagrangian in section \ref{sec:lagtran} using $\hat\mu,\hat\nu \ldots$ world indices in the full $D$ dimensions, and furthermore to place hats on all $D$-dimensional fields (and gamma matrices). When needed, $D$-dimensional tangent-space indices will be written as $\hat a,\hat b,\ldots$. We then use world indices $\mu,\nu,\ldots$ and tangent-space indices $\alpha,\beta,\ldots$ in the $(D-\dim G)$-dimensional spacetime and world indices
$m,n,\ldots$ and tangent-space indices $a,b,\ldots$ in the group manifold $G$.  Thus $\hat\mu=(\mu,m)$ and $\hat a=(\alpha,a)$, etc.

The $D$-dimensional bosonic field equations for the bosonic string, including conformal anomaly term,
are given in the string frame by
%%%%%
\bea
  \hat R - 4 (\del\Phi)^2 + 4 \hat\square\Phi -\fft1{12}
    \hat H^2 -\fft12 m^2 &=& 0\,,\label{phieqn}\\
  \hat R_{\hat\mu\hat\nu} + 2\hat\nabla_{\hat\mu} \hat\nabla_{\hat\nu}\Phi -\fft14 \hat H_{\hat\mu\hat\rho\hat\sigma}\,
   \hat H_{\hat\nu}{}^{\hat\rho\hat\sigma} &=&0 \,,\label{einsteqn}\\
  \hat\nabla_{\hat\mu}\,\Big(e^{-2\Phi}\, \hat H^{\hat\mu\hat\nu\hat\rho}\Big) &=&0\,.\label{Heqn}
\eea
%%%%%
We seek a ground-state solution whose metric is a direct sum of a $(D-\dim G)$-dimensional spacetime of maximal symmetry (Minkowski, AdS or dS) times the bi-invariant metric on the group manifold $G$:
%%%%%
\be
  d\hat s^2 = g_{\mu\nu}\, dx^\mu dx^\nu + g_{mn}\, dy^m dy^n\,.
\ee
%%%%%
The dilaton will be assumed to be constant, and taken, without material loss of generality, to vanish. The components of the 3-form $\hat M_{\hat\mu\hat\nu\hat\rho}$ will also be assumed to vanish except those lying entirely in the group-manifold, and for these we can take
%%%%%
\be
  \hat H_{mnp}= - c\, f_{mnp}\,.\label{Hans}
\ee
%%%%%
The choice of sign is arbitrary, as far as the bosonic equations of motion are concerned.  Our choice of the negative sign is for consistency with the pseudo-supersymmetry; see later. Here $f_{mnp}$ is constructed from the structure constants $f_{abc}$ using the vielbein $K_R^a$ in the obvious way:
%%%%%
\be
  f_{mnp}= K_{R\, m}^a\, K_{R\, n}^b\,K_{R\, p}^c\, f_{abc}\,.
\ee
%%%%%
It follows that we shall have
%%%%%
\be
  \hat H_{mn}^2 = c^2\, C_A\, g_{mn}\,,\qquad \hat H^2 = c^2\, C_A\, \dim G\,.
\ee
%%%%%

Plugging the ansatz into the dilaton field equation \eqref{phieqn} implies that we can take $\hat\phi=0$ if $m$ is given by
%%%%%
\be
  m^2= \fft13 c^2\, C_A\, \dim G\,.\label{mrel}
\ee
%%%%%
The $\hat R_{mn}$ components of the $\hat R_{MN}$ equation \eqref{einsteqn} then imply
%%%%%
\be
  R_{mn}= \fft14 c^2\, C_A\,g_{mn}\,,
\ee
%%%%%
which is precisely satisfied if the metric $g_{mn}$ on $G$ is taken to be the one considered in section \ref{sec:conventions}. The $\hat H_{\hat\mu\hat\nu\hat\rho}$ equation of motion \eqref{Heqn} is satisfied identically.  Since the equation for the mixed components $\hat R_{\mu n}$ of the Einstein equation is satisfied trivially, this leaves only the lower-dimensional spacetime components $\hat R_{\mu\nu}$ of the Einstein equation (\ref{einsteqn}), and this gives
%%%%%
\be
  R_{\mu\nu}=0\,.
\ee
%%%%%
Thus we have proved that we indeed have a Minkowski$\times G$ vacuum solution with $\hat\phi=0$ and $\hat H_{mnp}$ given by \eqref{Hans}, provided that the coefficient $m^2$ of the anomaly term is given by
\eqref{mrel} and that the metric $g_{mn}$ on the group manifold $G$ is chosen as described in section \ref{sec:conventions}.

An identical conclusion arises from the consistent truncation outlined in the last subsection. It is straightforward to see that a Minkowski space with a constant dilaton is captures by $F_{ABC}=0$ and $F_A=0$. Hence, solving the field equations for the product space Minkowski$\times G$ boils down to solving the field equations
\begin{equation}\label{eqn:DFTeom}
  \mathcal{R}_{AB} = 0 \qquad \text{and} \qquad \mathcal{R} - \frac{m^2}2 = 0
\end{equation}
in the internal space \cite{Hohm:2010pp}. Here $\mathcal{R}_{AB}$ denotes the generalised Ricci tensor and $\mathcal{R}$ is the generalised Ricci scalar . Both admit a very simple expressions for the generalised Scherk-Schwarz truncation we are concerned with
\begin{align}
  \mathcal{R}_{AB} &= 8 P_{(A}{}^C \Pb_{B)}{}^D \left( F_{CEG} F_{DFH} P^{EF} \Pb^{GH} + F_{CDE} F_F P^{EF} \right)\,, \\
  \mathcal{R} &=  P^{AB} P^{CD} \left( \Pb^{EF} + \frac13 P^{EF} \right) F_{ACE} F_{BDF} - 2 P^{AB} F_A F_B\,.
\end{align}
According to \eqref{eqn:FsforG}, only constructions of $F_{ABC}$ with exclusively $P^{AB}$ or $\Pb^{AB}$ give non-vanishing contributions. For this observation, we immediately see that $\mathcal{R}_{AB}$ vanishes (remember $F_A=0$) as expected. In the same vein, we obtain
\begin{equation}
  \mathcal{R} = \frac16 c^2 f_{abc} f^{abc} =  \frac16 c^2 C_A \dim G\,,
\end{equation}
and therefore recover \eqref{mrel} from the second equation in \eqref{eqn:DFTeom}.

\section{Pseudo-supersymmetry of the Minkowski\texorpdfstring{$\times G$}{xG} vacuum}
To check if this background at least partially preserves pseudo-supersymmetry, we need to plug it into the fermionic pseudo-supersymmetry transformation rules, to see whether $\delta\hat\psi_M$ and $\delta\hat\lambda$ vanish for some subset of the parameters $\hat\ep$.  The calculations can be set up along the same lines as those described in \cite{Duff:1986ya} for compactifications of $d=11$ supergravity. In particular, it will involve decomposing the spinors of the $D$-dimensional spacetime into tensor products of spinors in the $(D-\dim G)$-dimensional spacetime and spinors on the group manifold $G$. See appendix \ref{sec:Dirac} for a summary of how the Dirac matrices may be decomposed in the various cases of even or odd-dimensional spacetime and internal space.

As a preliminary check, consider the dilatino transformation rule in \eqref{eqn:SUSYtrafo1}, together with the conformal anomaly contribution in \eqref{conftrans}. In the background we are considering, with $\Phi=0$ 
and the 3-form given by \eqref{Hans}, we shall have
%%%%%
\be
  \delta\hat\lambda = \fft1{12} \, 
    c\, f_{abc}\, \hat\Gamma^{abc} \, \hat\ep + 
  \fft{\im}{2\sqrt2}\, \hat\ep\,.
\ee
%%%%%
We shall consider the case $\beta=-1$, where, as we discussed before, the conformal anomaly extension is simpler. Using
%%%%%
\be
  \hat\Gamma^{abc}\, \hat\Gamma_{def}= \hat\Gamma^{abc}{}_{def} +
    9 \hat\Gamma^{[ab}{}_{[de}\,\delta^{c]}_{f]} -
    18\hat\Gamma^{[a}{}_{[d}\, \delta^{bc]}_{ef]} -6\delta^{abc}_{def}\,,
\ee
%%%%%
it follows that if we define 
%%%%%
\be
\hat Q\equiv \fft1{6}\, f_{abc}\, \hat\Gamma^{abc}\,,
\ee
%%%%%
then
%%%%%
\be
  \hat Q^2 = -\fft1{6}\, f_{abc}\, f^{abc} = -\fft{C_A\, \dim G}{6}\label{Qsq}
\ee
%%%%%%
times the identity matrix.  By the Cayley-Hamilton theorem, and noting that ${\rm tr\,} \hat Q=0$, this means that $\hat Q$ has the eigenvalues
%%%%%
\be
  \pm \im \sqrt{\fft{C_A\, \dim G}{6}}\,,
\ee
%%%%%
with equal numbers of each. Thus with $m$ given, from \eqref{mrel}, by
%%%%%
\be
  m = c\, \sqrt{\fft{c_A\, \dim G}{3}}\,,\label{mval}
\ee
%%%%
we see that if $\hat\ep$ is any of the eigenvectors with eigenvalue $-\im \sqrt{\fft{C_A\, \dim G}{6}}$, 
we shall get $\delta\hat\lambda=0$.  The dilatino transformations suggest therefore that the Minkowski$\times G$ background preserves one half of the pseudo-supersymmetry.  

To confirm this, we now turn to the gravitino transformation rule. Assuming again that $\beta=-1$ we have, from \eqref{eqn:SUSYtrafo1},
%%%%%
\be
  \delta\hat\psi_{\hat\mu} = \hat D_{\hat\mu}\hat\ep
     -\fft18 \hat H_{\hat\mu\hat\nu\hat\rho}\, \hat\Gamma^{\hat\nu\hat\rho}\,.
\ee
%%%%%
In the internal group manifold directions we have
%%%%%
\bea
  \delta\hat\psi_m &=& \del_m\,\hat\ep + \fft14 (\omega_{ab})_m\, \hat\Gamma^{ab}
    \,\hat\ep + \fft{c}{8} f_{mnp}\, \hat\Gamma^{np}\,\hat\ep\,,\nn\\
  &=& \del_m\, \hat\ep\,,
\eea
%%%%%
after using the expression \eqref{spincon} for the spin connection on the group manifold. Finally, in the Minkowski spacetime directions we have
%%%%%
\be
  \delta\hat\psi_\mu = \del_\mu\, \hat\ep\,.
\ee
%%%%%
Thus, we see that the pseudo-supersymmetry variations of both the dilatino and the gravitino vanish provided that $\hat\ep$ is an eigenstate of $\hat Q$ with eigenvalue $-\im \sqrt{\fft{C_A\, \dim G}{6}}$, and that
$\hat\ep$ is independent of all the coordinates.

Again, we rederive this result using the relation between generalised geometry and pseudo-supersymmetry established in section~\ref{sec:genpseudo} where we consider the transformation of the gravitino first. Combining \eqref{eqn:susyvarGG}, \eqref{eqn:defnabla}, \eqref{eqn:formomega} and \eqref{eqn:FsforG} yields
\begin{equation}
  \delta \hat\psi_{\bar a} = \nabla_{\bar a}^{(-)} \hat\ep = \sqrt{2} \DD_{\bar a} \hat\ep + \frac1{\sqrt{2}} \omega_{\bar a b c} \hat\Gamma^{bc} \hat\ep = k_{R \bar a}^m \partial_m \hat\ep = 0\,,
\end{equation}
which tells us that the spinor $\hat\ep$ has to be constant. In the same vein we compute the variation of the generalised dilatino. It consists of two contributions: First, we evaluate
\begin{equation}\label{eqn:vard1}
  \delta \hat\rho = \hat\Gamma^a \nabla_a^{(+)} \hat\ep = \frac1{\sqrt{2}} \omega_{abc} \hat\Gamma^a \hat\Gamma^{bc} \hat\ep = \frac{1}{12} c f_{abc} \hat\Gamma^{abc} \hat\ep = \frac12 c \hat Q \hat\ep\,,
\end{equation}
where we take into account that partial derivatives on $\hat\ep$ have to vanish. Second, we include the conformal anomaly term, which alters the transformation of the generalised dilatino according to
\begin{equation}\label{eqn:vard2}
  \delta_{\rm extra} \rho = \frac{\im}{2\sqrt2} m \hat\ep
\end{equation}
for $\beta=-1$. Together, \eqref{eqn:vard1} and \eqref{eqn:vard2} yield
\begin{equation}
  \hat Q \hat\ep = - \frac{\im}{\sqrt{2}} m \hat \ep\,,
\end{equation}
which leads to the same result as already discussed above.

One may also consider more general vacuum solutions of the form (Minkowski)$_{D-\dim G} \times G_s$, where $G_s$ is a group manifold endowed with a ``squashed'' metric that, while still being invariant under the left action of the group $G_L$, is no longer invariant under the right action of the full group $G_R$.  We have looked at examples where $G$ is taken to be $SU(3)$ or $SO(5)$, and although these can indeed give rise to squashed solutions, we find that there is non surviving pseudo-supersymmetry in these backgrounds. The details of the $SO(5)$ example are described in appendix \ref{app:squashed}.

\section*{Acknowledgments}
We are grateful to Daniel Butter, Hong L\"u and Aritra Saha for useful discussions. This work is supported in part by DOE grant DE-FG02-13ER42020.

\appendix
\section{Spinors in \texorpdfstring{$D$}{D} Dimensions}\label{app:Spinors}
Here we reproduce a table from \cite{Lu:2011ku}, showing the types of spinor, and the corresponding values of the constant $\beta$, that can arise in each dimension.
\bigskip
\begin{table}[!ht]\centering
  \begin{tabular}{|c|c|c|c|c|c|c||c|c|}\hline
     $D$ mod 8 & $C\Gamma^{(0)}$ & $C\Gamma^{(1)}$ & $C\Gamma^{(2)}$ &
    $C\Gamma^{(3)}$ & $C\Gamma^{(4)}$ & $C\Gamma^{(5)}$ & Spinor &$\beta$\\
    \hline\hline
    0 & S & S & A & A & S & S & M & $+1$ \\
      & S & A & A & S & S & A & S-M & $-1$ \\ \hline
    1 & S & S & A & A & S & S & M & $+1$ \\ \hline
    2 & S & S & A & A & S & S & M & $+1$ \\
      & A & S & S & A & A & S & M & $-1$ \\ \hline
    3 & A & S & S & A & A & S & M & $-1$ \\ \hline
    4 & A & S & S & A & A & S & M & $-1$ \\
      & A & A & S & S & A & A & S-M & $+1$ \\ \hline
    5 & A & A & S & S & A & A & S-M & $+1$ \\ \hline
    6 & A & A & S & S & A & A & S-M & $+1$ \\
      & S & A & A & S & S & A & S-M & $-1$ \\ \hline
    7 & S & A & A & S & S & A & S-M & $-1$ \\ \hline
  \end{tabular}
  \caption{$\Gamma$-matrix symmetries and spinor representations
  in diverse dimensions. S denotes symmetric, A denotes antisymmetric, 
  M denotes Majorana and S-M denotes symplectic Majorana.}\label{tab:Spinors}
\end{table}

\section{Decomposition of Dirac Matrices}\label{sec:Dirac}
In a Kaluza-Klein reduction we need to write the higher-dimensional Dirac matrices $\hat \Gamma_A$ in terms of tensor products of lower-dimensional spacetime Dirac matrices $\gamma_\alpha$ and internal space Dirac matrices $\Gamma_a$. The way this works depends upon whether the various space(time)s are even-dimensional or odd-dimensional. A table of how the decompositions may be made is given in appendix A of \cite{Lu:1998nu}:
%%%%%
\begin{align}
  \hbox{(even,odd)}:&& \qquad \hat\Gamma_\alpha&=\gamma_\alpha\otimes\oneone \,,
  \qquad & \hat\Gamma_a&=\gamma_*\otimes\Gamma_a\,,\nn\\
  \hbox{(odd,even)}:&&\qquad \hat\Gamma_\alpha&=\gamma_\alpha\otimes\Gamma_*\,,
  \qquad & \hat\Gamma_a&=\oneone\otimes\Gamma_a\,,\nn\\
  \hbox{(even,even)}:&&\qquad \hat\Gamma_\alpha&=\gamma_\alpha\otimes\oneone\,,
  \qquad & \hat\Gamma_a&=\gamma_*\otimes\Gamma_a\,,\nn\\
  \hbox{or} &&\qquad \hat\Gamma_\alpha&=\gamma_\alpha\otimes\Gamma_*\,,
  \qquad & \hat\Gamma_a&=\oneone\otimes\Gamma_a\,,\nn\\
  \hbox{(odd,odd)}:&&\qquad \hat\Gamma_\alpha&=\sigma_1\otimes\gamma_\alpha\otimes\oneone\,,
  \qquad & \hat\Gamma_a&=\sigma_2\otimes\oneone\otimes\Gamma_a\,,
\end{align}
%%%%%
where the first entry in the pair enclosed in parentheses indicates whether the lower-dimensional spacetime is even or odd-dimensional, and the second entry indicates whether the internal space is even or odd-dimensional.
$\gamma_*$ denotes the chirality operator in even-dimensional lower-dimensional spacetimes, and $\Gamma_*$ denotes the chirality operator in even-dimensional internal spaces (with $\gamma_*^2=+1$ and $\Gamma_*^2=+1$).  In the (odd,odd) case the extra factor involving the Pauli matrices $\sigma_1$ and $\sigma_2$ ensures that the $\hat\Gamma_A$ matrices obey the Clifford algebra.  They are needed because the Dirac matrices $\hat\Gamma_A$ in this case are twice the size of the tensor products of the lower-dimensional and the internal Dirac matrices.

\section{Squashed Group Manifold Solutions}\label{app:squashed}
It is well known that any compact semi-simple group manifold other than $SU(2)$ or $SO(3)$ admits at least one additional, inequivalent, Einstein metric, over and above the standard bi-invariant metric. This raises the possibility that there might exist Minkowski$\times G$ vacua in which the metric on the group manifold $G$ is not the bi-invariant one. Such solutions would not necessarily involve a squashed Einstein metric on $G$, since the form of the 3-form field strength $H_{mnp}$ in the squashed vacuum may also change. One approach to looking for such squashed solutions is to consider families of squashed metrics on $G$, with an associated deformation of the 3-form field.  The families of metrics in question here will be homogeneous, invariant still under the left-acting copy of $G$, but no longer invariant under the full right action of $G$.  Such metrics can be obtained by rescaling the left-invariant vielbeins by constant factors.  A detailed discussion of the construction of squashed Einstein metrics using this procedure can be found, for example, in \cite{Gibbons:2009dx}.

One can look for squashed vacuum solutions on a case by case basis. We have checked two examples, one being a family of squashed metrics on the $SU(3)$ group manifold and the other a family of squashed metrics on the $SO(5)$ group manifold. In neither case do we find any squashed vacua in the bosonic string for which the squashed metric on the group manifold is of positive-definite signature. For the case of $SO(5)$ we do find one squashed example for which the metric has Lorentzian signature (that is, $(1,9)$ signature).  Since this may be of some interest, we shall present some details below.

We make define left-invariant 1-forms $L_{IJ}$ for $SO(5)$, with $I$ and $J$ ranging over 1 to 5, and $L_{IJ}=-L_{JI}$, obeying the exterior algebra
%%%%%
\be
  dL_{IJ}= L_{IK}\wedge L_{KJ}\,.
\ee
%%%%%
Following \cite{Gibbons:2009dx} we split the indices into $I=(1,2,i)$, and here we define the ``unsquashed'' vielbein
%%%%%
\be
  \bar e_i=L_{1i}\,,\quad \bar e_{i+3}= L_{2i}\,,\quad
  \bar e_7=L_{34}\,,\quad \bar e_8=L_{35}\,,\quad e_9=L_{45}\,,
  \quad \bar e_{10}=L_{12}\,.
\ee
%%%%%
When we need to assign specific numerical values to tangent-space indices it is more convenient to put the index downstairs. We consider metrics
%%%%%
\be
  ds_{10}^2 = x_1\,(\bar e_1^2+\bar e_2^2 +\bar e_3^2)  
    + x_2\, (\bar e_4^2+\bar e_5^2 +\bar e_6^2) +
    x_3\, (\bar e_7^2+\bar e_8^2 +\bar e_9^2) +x_4\, \bar e_{10}^2\,,
  \label{squashedso5}
\ee
%%%%%
where $(x_1,x_2,x_3,x_4)$ are constants. Correspondingly, we have a vielbein
%%%%%
\be
  e_1=\sqrt{x_1}\, \bar e_1\,,\qquad e_2=\sqrt{x_1}\, \bar e_2\,,\ldots
  \,, e_9=\sqrt{x_3}\, \bar e_9\,,\qquad
  e_{10}=\sqrt{x_4}\, \bar e_{10}\,.
\ee
%%%%%
Thus we have a four-parameter family of homogeneous metrics on $SO(5)$ which are invariant under the left action of the full $SO(5)$ group but invariant only under an $SO(3)$ subgroup of right-acting transformations.  As was discussed in \cite{Gibbons:2009dx}, there are three inequivalent Einstein metrics in this family, corresponding to (up to overall scale)
%%%%%
\be
  (x_1,x_2,x_3,x_4)=\quad (1,1,1,1)\,,\qquad
  (14,14,4,19)\,,\qquad (1,2,1,2)\,.
\ee
%%%%%
The first of these is the standard bi-invariant metric.

In order to obtain a solution of the bosonic string of the form (Minkowski)$\times G_{squashed}$, we need also to construct a 3-form $G_\3$ that is closed and also co-closed (i.e.~the 3-form must be harmonic).  In the case of the bi-invariant vacuum, we just used a constant multiple of the structure constants $f_{abc}$. In fact we could write
%%%%%
\be
  G_\3 = \fft13 d\bar e^a\wedge \bar e^a = \fft16 f_{abc}\, \bar e^a\wedge
  \bar e^b\wedge \bar e^c\,.\label{G3}
\ee
%%%%%
This is manifestly closed, and one can easily verify that it is also  co-closed in the bi-invariant metric.  

There must always exist an harmonic 3-form regardless of whether the metric is bi-invariant or squashed, since the topological number $b_3$ (the third Betti number) is equal to 1 regardless of the metric. One way to construct the required harmonic 3-form is by a brute-force Mathematica calculation, starting with a general 3-form
%%%%%
\be
  G_\3=\fft16 G_{abc}\, e^a\wedge e^b\wedge e^c 
\ee
%%%%%
and solving for the (constant) components $G_{abc}$ such that $dG_\3=0 =d{*G_\3}$.  In fact we find that the harmonic 3-form is exactly the same as the one constructed in eqn (\ref{G3}) (i.e.~still written using the bi-invariant vielbein $\bar e^a$).  Of course when one calculates the Hodge dual and $d{*G_\3}$ (or equivalently, the divergence $\nabla^a\, G_{abc}$), the fact that the metric is squashed enters in the calculation.

Substituting $\hat H_{mnp}= \pm c G_{mnp}$ and the direct sum of the Minkowski metric and the squashed $SO(5)$ metric (\ref{squashedso5}) into the equations of motion for the bosonic string (with dilaton set to zero), we find two inequivalent solutions.  Up to scaling, they are %%%%%
\be
  (x_1,x_2,x_3,x_4)=(1,1,1,1)\,,\qquad c=\fft13\,,\qquad m^2=20\,,
\ee
%%%%%
which is the bi-invariant solution of the kind we found earlier for a general group $G$, and 
%%%%%
\be
  (x_1,x_2,x_3,x_4)= (1,1,3,-3)\,,\qquad c=1\,,\qquad
    m^2=\fft{56}{3}\,.\label{squashedso5sol}
\ee
%%%%%
We see from (\ref{squashedso5}) that the squashed metric on $SO(5)$ in this solution has Lorentzian $(1,9)$ signature, because $x_4$ is negative.  

It was noted in \cite{Gibbons:2009dx} that although various examples of squashed group manifolds were checked and many squashed Einstein metrics were found, all them had either Euclidean signature or else more than one timelike direction.  Our squashed $SO(5)$ bosonic string vacuum \eqref{squashedso5sol} thus provides a first example of a Lorentzian signature group manifold metric arising as a solution in a theory of physical interest.We can, of course, take the flat directions in the vacuum solution to be Euclidean space rather than Minkowski spacetime in this case, so that the signature of the entire higher-dimensional bosonic string spacetime will be $(1,D-1)$.

\bibliography{literature}
   
\bibliographystyle{JHEP}
\end{document}